\documentstyle[twocolumn,aps,prl,epsf]{revtex}
\begin{document} 

\title{Direct Mott Insulator-to-Superfluid Transition in the 
Presence of Disorder}
\author{Ferenc P\'azm\'andi and Gergely T. Zim\'anyi}
\address{Physics Department, University of California, Davis, CA 95616}
\address{\mbox{ }}

\address{\parbox{14cm}{\rm \mbox{ }\mbox{ }
We introduce a new renormalization group theory to examine 
the quantum phase transitions upon exiting
the insulating phase of a disordered, strongly interacting boson
system. For weak disorder we find a direct transition 
from this Mott insulator to the Superfluid phase.
In $d > 4$ a finite region around the
particle-hole symmetric point supports this direct transition, whereas for
$2\le d<4$ perturbative arguments suggest that the direct transition 
survives only precisely at commensurate filling.
For strong disorder the renormalization trajectories  pass
next to two fixed points, describing a pair of distinct transitions; first from
the Mott insulator to the Bose glass, 
and then from the Bose glass to the Superfluid. 
The latter fixed point possesses statistical particle-hole symmetry
and a dynamical exponent $z$, equal to the dimension $d$.
}}
\address{\mbox{ }}
\address{\parbox{14cm}{\rm \mbox{ }\mbox{ }
}}
\address{\mbox{ }}
\maketitle

\narrowtext

The field of quantum phase transitions in disordered systems experienced
remarkable growth recently. Its experimental relevance comprises the
superconductor-insulator transition in thin films\cite{hTc}, $^4$He 
in disordered media \cite{reppy}, as well as Quantum Hall systems,
and many types of disordered magnets\cite{sachdev}. 
Quantum criticality may lie at the heart of the anomalous normal 
state properties of high $T_c$ superconductors, and it can 
describe the characteristics of vortex systems with correlated disorder
\cite{vinokur}. The intensely studied localization transition of interacting,
disordered bosons has served as a useful paradigm for quantum phase transitions.
Most of the originally proposed theoretical picture\cite{fisher}
was subsequently confirmed numerically\cite{young}.
However a key issue remains controversial. Analytic theories suggest, 
that the localized, ``Bose Glass" phase intervenes everywhere between the interaction
driven (Mott) insulator and the superfluid\cite{fisher}.  
However, none of the numerical studies of the problem
\cite{krauth,rieger} found the Bose glass in $2d$ at 
commensurate densities. Experimental tests in Josephson junction arrays also 
analyzed the critical behaviour\cite{mooij}. In these systems charge neutrality
requires commensurate boson densities. Also, some disorder is unavoidably
present, as proven by the analysis of the transition in finite magnetic fields
in the same samples. Despite this disorder, in the vicinity of the zero 
field transition the temperature dependence of the resistivity and other
quantities again do not indicate the presence of a Bose glass phase. 

In this letter we present a novel renormalization group analysis of the
problem. We find that in dimensions $d\ge 2$ a {\it direct} insulator-superfluid
transition takes place for weak disorder in these systems. Furthermore the
dynamical critical exponent $z=d$ at the Bose Glass - Superfluid transition,
where a statistical version of particle-hole symmetry develops.

We consider the Hamiltonian:
\begin{equation}
{\sl H} = -\sum_{i,j} J_{ij} a_{i}^{\dagger}a_{j} - \sum_{i} \mu _{i} 
\hat n_i +U\sum_i \hat n_i(\hat n_i-1)
\label{H}
\end{equation}
where $a_{i}$ ( $a_{i}^{\dagger}$) annihilate (create) a boson
at site $i$ and $\hat n_i=a_{i}^{\dagger}a_{i}$ is the number operator, and
$\mu_{i}=\mu+\epsilon_{i}$, where
the chemical potential $\mu$ controls the density of the bosons, 
and $\epsilon_{i}$ is a random
site energy, with distribution $P(\epsilon)$ over the {\it finite} support 
$[-D, D]$. 

For clean systems the phase diagram consist of a
Superfluid phase (SF) and lobe-like Mott insulating phases (MI)
with integer boson densities \cite{fisher}.
In the presence of disorder some site energies fall within the clean gap,
shrinking the Mott lobes. It was proposed that a new, Bose Glass
(BG) phase appears between the superfluid and the insulator, characterized by the
absence of a SF order parameter, and a Mott gap. 
The excitations are presumed localized,
with a finite density of states (DOS) at low energies and thus 
with a nonvanishing compressibility\cite{fisher}.

We first present a mean field treatment of the problem by
chosing $J_{ij}=J/N$, connecting all sites. In this formulation
the asymptotics of the distribution of the site disorder plays a crucial
role. It has recently been shown that for $P(\epsilon)\sim (D-|\epsilon|)^\alpha$,
with $\alpha>0$, there is indeed a localization transition on the mean field level
\cite{us}. A physical motivation for such a choice is that
spatial fluctuations of the field are likely to renormalize the customary 
uniform distribution to a smooth function in finite dimensions\cite{note1}.
As this does not happen by itself on the mean-field level, 
we instead build this feature in from the start
by choosing a smooth unrenormalized distribution. 
We find that the critical behaviour at the MI-SF and the BG-SF
transitions is {\it independent} of the details of the site-energy distribution.

The evaluation of the free energy is achieved by performing a Hubbard-Stratonovich transformation on the kinetic term.
In the $N \rightarrow \infty$ limit the dynamic fluctuations
are suppressed, thus the saddle point approximation becomes exact,
yielding for the free energy
\begin{eqnarray}
\beta f&=&-N\beta J{|m|^2}+
     \sum_i^N \ln Tr \exp[-\beta S_i(m)]\\
S_i(m)&=&-\mu_i \hat n_{i} +U \hat n_{i} (\hat n_{i}-1)- J 
(m^* a_{i}+m a_{i}^{\dagger})~~,
\label{f}
\end{eqnarray}
where $m$ takes its saddle-point value. $m$ is proportional
to the superfluid order parameter so it is zero
in both the MI and BG phases. The MI phase is distinguished from the BG
by the presence of a gap in its spectrum. The MI-BG transition is driven by the 
local collapse of this gap at some sites, which induces singular
contributions to $f$. While this is the dominant scenario at small $J$,
for stronger kinetic couplings a direct MI-SF transition
is possible: long range fluctuations might generate a nonzero
value for $m$ before local instabilities would take place. In this
case the free energy can be expanded in powers of $m$ and a
Landau-type action describes the MI-SF transition yielding the
usual mean-field exponents.

The advantage of employing a smooth distribution function of the disorder
becomes apparent here. For the uniform distribution the MI-BG transition 
is absent for $J>0$\cite{fisher}, whereas it is manifestly present at 
some finite $J$ for positive $\alpha$'s \cite{us}.

We give credence to the above physical picture by developing
a new renormalization group (RG) scheme, similar to that of Ref.\cite{danfish}.
As will be shown below, the soft core problem always scales to large interaction
strengths, therefore we first present the procedure for hard core bosons
and will discuss the effect of finite $U$ later.
Minimizing the free energy yields the saddle-point equation at
$T \rightarrow 0$
\begin{equation}
m=m~{J\over N}\sum_i^N 1\bigg{/}\sqrt{\mu_i^2+J^2m^2}~~.
\label{SP}
\end{equation}

This sum is unfortunately plagued with singular terms with $\mu_i \sim 0$.
The key concept of our RG technique is to avoid these singular terms by 
performing the sum step by step, integrating out those {\it sites} which 
have the highest excitation energies: $E = \max_{i}|\mu_{i}|$. 
The contribution of these sites is then exactly
absorbed into the renormalization of the parameters as:
\begin{eqnarray}
dm/dx&=&m \Bigl( 1-J\big/\sqrt{E^2+J^2m^2}\Bigr)\cr
dJ/dx &=& J\Bigl( \tilde{z} - 1 + J\big/\sqrt{E^2+J^2m^2}\Bigr)\cr
dD/dx &=& D \Bigl(\tilde{z} -1/(\alpha +1) \Bigr) ~~\label{4}\cr
dE/dx &=& \tilde{z}E - D/(\alpha +1) ~~\label{5}.
\end{eqnarray}
Here $dx=-d\rm{ln}N$ parametrizes the 
change in the number of sites, and the dynamic critical exponent $\tilde{z}$
relates this change to the rescaling the unit of energy $\Omega$:
$d\Omega/\Omega = \tilde{z} dN/N$. 
These equations apply in the asymptotic region, 
where $P(\epsilon) \sim (D+\epsilon)^\alpha$. This region 
is reached by integrating out the high energy sites with $\epsilon_{i}>0$
in one step. Keeping the {\it total} bandwidth fixed, $E = 1 $, requires 
$\tilde{z}=D/(\alpha +1)$.

The flow trajectories of the scaling equations are displayed in Fig.1.
for the case of $m=0$.
We find an {\it attractive} critical fixed point at $D=J=0$.
In the absence of disorder and hopping this point is naturally
identified as the Mott Insulator. The {\it critical} fixed point 
at $D=0$, $J=1$ separates from the MI phase a region with runaway 
trajectories towards large $J$ and vanishing disorder $D$. These 
trajectories are regularized by 
inducing finite values of $m$: thus they characterize a superfluid. 
This identification of a {\it direct} Mott-Superfluid transition
for weak disorder is the central result of our paper.
 
\begin{figure}
\epsfxsize=3.0in
\epsfysize=2.25in
\epsffile{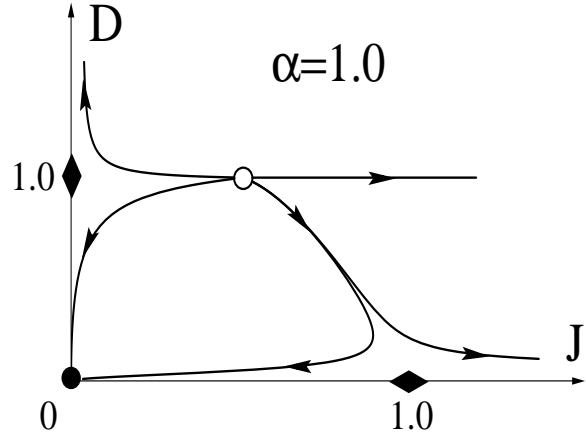}
\vskip 0.2cm
\caption{The renormalization group trajectories for $\alpha=1$.}
\end{figure}

There is a {\it critical} point at $D=1$, $J=0$, which sends trajectories
towards large $D$ and $J=0$, thus describing the MI-BG transition.
As $D$ increases, at the sites with the lowest local energy $\mu_i<0$, the cost 
of adding a particle, $|E-D|$ eventually becomes equal to $E$. Since $E$ is 
kept fixed at $1$, this happens at $D=2$, where half of the sites becomes empty. 
Subsequent RG steps should involve the elimination of the highest energy
particle {\it and} hole excitations, from the two edges of the distribution
simultaneously. From now on the trajectories remain in the $D=2$ plane,
since the width of the distribution (now determined by $E=D/2$) is kept fixed.
However the {\it shape of the distribution} of the site energies renormalizes.
Even if the original distribution is asymmetric, this procedure stretches
the narrow section around the zero of energy up to the original bandwidth.
Obviously the fixed point distribution is {\it uniform}, with half of the sites
empty, representing a {\it statistical particle-hole symmetry}, 
even though the original model did not possess this symmetry\cite{phsymm}.

The fixed point distribution is thus characterized by $\alpha = 0$, as well as
$E=1$, $\tilde{z}=1$, yielding
\begin{eqnarray}
dm/dx&=&m (1-J) ~+ {\cal O}(m^3) ~~\cr
dJ/dx &=& J^2 ~+ {\cal O}(m^2)
\end{eqnarray}
The fixed point is at $J=0$, $D=2$, $m=0$, which is naturally identified
as the BG-SF critical fixed point.
Combining the equations for $m$ and $J$ we get 
$dm/dJ= m(1/J^2-1/J)$, yielding $m \sim (1/J) \exp(-1/J)$\cite{us}. 

The following picture emerges for the flow trajectories. For disorder 
values $D>1$ the system first scales close to the MI-BG fixed point
at $D=1$, $J=0$, which is characterized by the exponents, such as
$\tilde{z}=1/(\alpha +1)$. But the flows continue towards 
the BG-SF fixed point at $D=2$, $J=0$. The corresponding critical 
behaviour becomes {\it independent} of $\alpha$ as the disorder
distribution scales to a uniform function ($\alpha=0$).
In particular, this restores the statistical particle-hole symmetry at the 
BG-SF fixed point, with $\tilde z=1$.
Here let us recall that our infinite-range hopping model is equivalent 
to the mean-field treatment of a finite-dimensional system.                                                                                                    In a finite dimension $d$ the dynamical exponent $z$ relates
the rescaling of the linear size, $L$, of the system to the energy-scale
change $\Omega'/\Omega=(L'/L)^z$. By writing $(N'/N)=(L'/L)^d$, 
from $\tilde z=1$ we obtain $z=d$ for the BG-SF transition.
This result was first derived from the scaling of the compressibility
in Ref.\cite{fisher}, but was recently debated \cite{weichman}.
Finally, the special, {\it multicritical} fixed point at $D=1$, 
$J=\alpha/(\alpha+1)$ will be analyzed in a subsequent publication.

\begin{figure}
\epsfxsize=3.0in
\epsfysize=2.25in
\epsffile{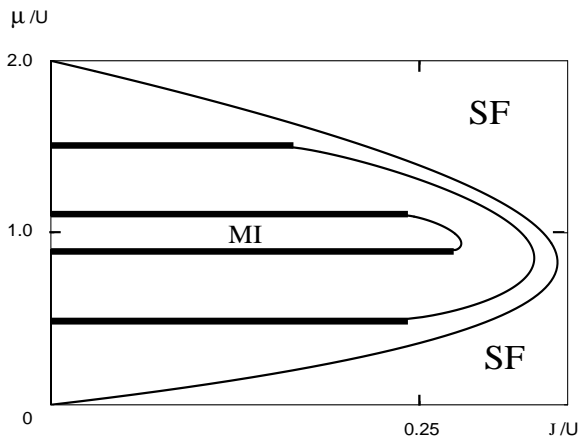}
\vskip 0.2cm
\caption{The phase boundary of the first Mott lobe for $\alpha=1$,
for disorder values $ D=0, ~0.5, ~0.9$. 
The thick line represents the Bose-Glass.}
\end{figure}

Now we return to the case of soft core bosons.
The novelty is that from the start particle and hole excitations
{\it at the same site} have to be considered simultaneously.
Formally this is implemented by expanding the saddle point equation in $m$,
and then replacing $1/|\mu_i|$ by 
$ (n_i/E_{h,i} + (n_{i}+1)/E_{p,i})$,
where $n_i$ is the number of bosons at site $i$ in the ground state.
Its value is determined from $2U(n_i -1)<\mu_i<2Un_i$. 
The excitation energy of a hole is $E_{h,i}=\mu_i-2U(n_i -1)$, 
and of a particle $E_{p,i}=2Un_i-\mu_i$. Finally, $U$ renormalizes, as
$dU/dx= \tilde{z}U$. Let us observe that
as long as the disorder does not scale to zero, $\tilde z = D/(\alpha +1)$ 
is positive, driving $U$ to infinity. Thus for the finite $D$ transitions
the trajectories indeed scale to $U=\infty$, thus we find the {\it same} fixed points, as in the hard core case. When the system scales towards small
disorder, $U$ increases, but saturates at a nonuniversal finite value. 
This stretches the hard core MI and the MI-SF fixed points into fixed lines.
Therefore the fixed point structure of the soft core case is equivalent
to that of the hard core case. 

We explicitly calculated the phase boundary of the Mott Insulator for 
soft core interactions
by requiring that the coefficient of $m$ on the {\it rhs.} of the appropriately
modified saddle point equation
equal 1. Since at the MI-BG phase boundary $\mu$ is either $D$ or $2U-D$, 
the straight separatrix at $D=1$ in the $D-J$ plane translates into 
straight sides of the Mott lobes, as shown in Fig.2.
It is worth noting that with increasing disorder the Mott lobe shrinks much faster
along the $\mu$ axis, than along the $J$ axis, in contrast to 
previous suggestions\cite{fisher,monien}.

Now we address the question of what happens on finite
dimensional lattices. First the pure case will be reviewed,
and then a small disorder turned on. We concentrate on
the regions around the tip of the Mott lobes, where the transition
happens at finite hopping $J$, so $D$ can be regarded as a
small parameter.

In the pure case there are two types
of transitions: exiting the MI by changing the chemical potential $\mu$
(generic case) and by increasing the hopping $J$ at the tip of 
the lobe (multicritical point)\cite{fisher}. The generic transition is
driven by particle or hole excitations. The dynamical exponent
is $z=2$ and the upper critical dimension is $d_c=2$. The gap
linearly disappears with $\delta \mu=\mu-\mu_c$, requiring $\nu z=1$,
and $\nu=1/2$ for any dimension $d\ge 1$. However the correlations
inside the ground state do not diverge upon approaching the transition.
On the other hand, the multicritical 
point possesses particle-hole symmetry, setting the value of $z$ to $1$: 
the transition is of the type of a $d+1$ dimensional $XY$ model, hence $d_c=3$.
The correlation length of the MI ground state does diverge upon approaching 
the transition.

Now we include site disorder $D$ in the model. The relevancy of the disorder
was extensively studied in past work. In classical models the 
perturbative Harris criterion is generally accepted to signal the 
{\it irrelevancy} of disorder\cite{harris}. In their non-perturbative study, 
Chayes et al. proved that if the transition
can be tuned by changing the disorder around a finite disorder fixed point,
then the correlation length exponent must obey $\nu > 2/d$\cite{chayes}.
This coincides with the Harris criterion for classical systems, but
is expected to apply for quantum transitions as well.
In the quantum case, however, perturbative considerations may yield further
bounds on exponents.
Imagine for instance of tuning the system to criticality. Here 
$\omega(k) \sim k^z$, by the definition of the dynamical critical exponent.
In a finite size system this yields a level spacing for the {\it first excited
states} of the clean Hamiltoninan $\delta \omega \sim L^{-z}$. 
On the other hand, disorder introduces a mixing between these states, of the
order of $V_{kk'} \sim L^{-d/2}$. Second order perturbation theory shows 
that the disorder - induced shift of the energy levels is much
less than the level spacing for $d \ge 2z$. Also, the perturbative
change in the ground state wavefunction remains small, thus disorder is
irrelevant. All of the above arguments suggest that 
at the generic transition, where $\nu=1/2$ and $z=2$, {\it weak} 
disorder is irrelevant for $d>d_c=4$. Thus the Bose-glass covers the sides 
of the Mott lobe only partially, leaving a finite region around the tip,
where a direct MI-SF transition occurs, as in mean field. 
On the other hand, for $d<4$ disorder becomes relevant, suggesting that 
the Bose-glass covers the sides of the Mott lobe completely.

At the tip the Chayes criterion gives $d_c=4$ again,
however the perturbative criterion yields $d_c=2$, since $z=1$. 
These two results may be reconciled by considering the following possibility.
For $D>0$ one could argue that the randomness hinders ordering
tendencies, thus the ordered superfluid phase should appear only at 
larger hoppings $J$. On the other hand, site - disorder is moving some 
energy levels {\it inside} the Mott gap. These sites may be able to support 
a superfluid even for hoppings smaller than $J_c$, the critical value
for the clean system. 
Thus it is not inconceivable that these two competing influences
cancel each other, and the phase boundary remains {\it independent} of
the disorder upto some finite $D$ value. In this case the Chayes et al. bound
does not apply, as its proof requires that the transition be driven by
tuning the disorder. According to the other inequality then, the
disorder is relevant only for $d < 2$. Remarkably, all numerical
studies, performed at the tip of the Mott lobes, find a {\it  direct
MI-SF transition}\cite{krauth,rieger}. In the recent 2$d$
work of Kisker and Rieger\cite{rieger} it was found that for $D < 0.4 J$ the 
$\nu$ exponent at the tip of the lobe approximately equaled its clean value of
$\nu_{3dXY}$, violating the Chayes et al. bound.
Correspondingly, the locus of the phase transition was
consistent with the clean value, allowing for the possibility of systematic errors.
Even more convincingly, the transition was characterized by genuinely 
new exponents around $D\approx 0.4$,
such as $z=1.4$, corresponding neither to the generic, nor to the 
$XY$ criticality. Such a new multicritical behaviour is indeed expected,
if a critical value of the disorder, $D_c$ exists, such that
for $D>D_c$ the disorder is relevant.


This scenario suggests that there can be a direct MI-SF
transition even for finite disorder at the tip of the Mott lobe, for $d\ge2$. 
One can rationalize this result by observing
that the correlation length diverges upon approaching the critical point 
from either side. One expects the disorder to be screened and smoothed
quite effectively by fluctuations of such large spatial extent.

We wish to end on cautionary notes. Obviously these same numerical results
can be in accord with the Chayes et al. bound, if the Bose glass phase
is extremely slim, or if it manifests itself only on length scales, exceeding
the biggest accesible system size. Such an anomalous behaviour 
however lacks a theoretical explanation to date. Additional subtleties
associated with applying the Chayes et al. criterion around multicritical
points were emphasized in Ref.\cite{singh}.

In sum, we studied interacting bosons in the presence of disorder on a lattice.
We constructed the phase diagram on the mean field level using an unconventional
renormalization group scheme. At weak disorder a direct Mott Insulator- Superfluid
transition takes place. We argued that this transition is present
for $d > 4$ in a finite region around the lobe-tips,
whereas it survives down to two dimensions at the tip of the Mott lobes.
Several numerical studies are consistent with this picture, as well as the 
limited experimental evidence on Josephson junction arrays. We proposed
a possible reconciliation of these results with the well-known bound
for $\nu$ of Chayes et al. At strong disorder, we found that
the Bose glass - Superfluid  transition
is characterized by statistical particle - hole symmetry
and the exponent relation $z=d$ on the mean field level.

We acknowledge useful discussions with R. Scalettar,
L. Chayes, T. Giamarchi, H. Rieger, 
S. Sachdev, R. Singh, P. Weichman, and P. Young. 
This research was supported by NSF DMR-95-28535.


\begin{references}
\bibitem{note1} Application of a version of our renormalization group
scheme in one dimension explicitly demonstrates this.
\end{references}
\end{document}